\documentclass[a4paper,fleqn,usenatbib]{mnras}

\usepackage[T1]{fontenc}
\usepackage{ae,aecompl}

\usepackage{graphicx}	
\usepackage{amsmath}	
\usepackage{amssymb}	
\numberwithin{equation}{section}
\DeclareMathOperator\erfinv{erfinv}
 
\usepackage{booktabs,fixltx2e}
\usepackage{verbatim}

\title[Measuring black hole spin with stellar orbits]{What stellar orbit is needed to measure the spin of the Galactic center black hole from astrometric data?}
\author[]{Idel Waisberg$^{1}$\thanks{E-mail: idelw@mpe.mpg.de}, Jason Dexter$^{1}$\thanks{jdexter@mpe.mpg.de}, Stefan Gillessen$^{1}$, Oliver Pfuhl$^{1}$, Frank Eisenhauer$^{1}$, \newauthor Phillip M. Plewa$^{1}$, Michi Baub\"ock$^{1}$, Alejandra Jimenez-Rosales$^{1}$, Maryam Habibi$^{1}$, \newauthor Thomas Ott$^{1}$, Sebastiano von Fellenberg$^{1}$, Feng Gao$^{1}$, Felix Widmann$^{1}$ \newauthor and Reinhard Genzel$^{1,2}$ \\
$^{1}$Max-Planck-Institut f\"{u}r Extraterrestrische Physik (MPE), Gie\ss enbachstra\ss e, 85741 Garching bei M\"{u}nchen, Germany\\
$^{2}$Physics Department, University of California Berkeley, CA 94720, USA \\
}

\date{Accepted XXX. Received YYY; in original form ZZZ}

\pubyear{2017}

\begin{document}
\label{firstpage}
\pagerange{\pageref{firstpage}--\pageref{lastpage}}
\maketitle

\begin{abstract}
Astrometric and spectroscopic monitoring of individual stars orbiting the supermassive black hole in the Galactic Center offer a promising way to detect general relativistic effects. While low-order effects are expected to be detected following the periastron passage of S2 in Spring 2018, detecting higher-order effects due to black hole spin will require the discovery of closer stars. In this paper, we set out to determine the requirements such a star would have to satisfy to allow the detection of black hole spin. We focus on the instrument GRAVITY, which saw first light in 2016 and which is expected to achieve astrometric accuracies $10-100 \mu$as. For an observing campaign with duration $T$ years, $N_{obs}$ total observations, astrometric precision $\sigma_x$ and normalized black hole spin $\chi$, we find that $a_{orb}(1-e^2)^{3/4} \lesssim 300 R_S \sqrt{\frac{T}{4 \text{years}}} \left(\frac{N_{obs}}{120}\right)^{0.25} \sqrt{\frac{10 \mu as}{\sigma_x}} \sqrt{\frac{\chi}{0.9}}$ is needed. For $\chi=0.9$ and a potential observing campaign with $\sigma_x = 10 \mu$as, 30 observations/year and duration 4-10 years, we expect $\sim 0.1$ star with $K<19$ satisfying this constraint based on the current knowledge about the stellar population in the central 1". We also propose a method through which GRAVITY could potentially measure radial velocities with precision $\sim 50$ km/s. If the astrometric precision can be maintained, adding radial velocity information increases the expected number of stars by roughly a factor of two. While we focus on GRAVITY, the results can also be scaled to parameters relevant for future extremely large telescopes.
\end{abstract}




\section{Introduction}
\label{sec:intro}

The orbits of short period stars in the central 1" (S-stars) of the Milky Way Galaxy provide the best current evidence for the existence of supermassive black holes. Currently $\approx 40$ orbits are known \citep{Gillessen17}, including that of the star S2, reaching $R \approx 1300 R_S$ from the black hole, where $R_S = 2GM/c^2$. Such orbital monitoring has led to strong constraints on the black hole mass and distance to the Galactic center \citep{Ghez08,Gillessen09,Boehle16,Gillessen17}.

The orbits are all currently compatible with Newtonian gravity. Lower order effects such as periastron advance and gravitational redshift are expected to be probed with the star S2 \citep{Jaroszynski98,Fragile00,Rubilar01,Weinberg05,Zucker06,Angelil10a,Hees17,Parsa17,Grould17} during or following its next closest approach in Spring 2018. However, Newtonian perturbations from a distribution of stars or remnants in the central region are very likely to dominate over higher order relativistic effects related to black hole spin for the currently known stars \citep[e.g.][]{Merritt10,Zhang17}. Detection of black hole spin from Lense-Thirring precession therefore requires the discovery and monitoring of closer stars. 

Several works have pointed to the possibility of using astrometric measurements of closer stars to constrain the black hole spin \citep[e.g.][]{Kraniotis07,Will08,Merritt10,Sadeghian11,Psaltis16}, but the technology to find such stars and achieve the required precision was lacking. This, however, has changed with the first light of the instrument GRAVITY at the Very Large Telescope Interferometer \citep[VLTI,][]{GRAVITY17}, one of whose main goals is to resolve the inner region around SgrA* at few mas resolution in search for closer stars, and to achieve $\sim 10-100 \mu$as astrometric precision in the monitoring of stellar orbits \citep{Eisenhauer11}. 

Recent investigations have explored possible constraints on the black hole spin of SgrA* using closer stars \citep{Zhang15,Yu16} assuming a combination of astrometric $1-30\mu$as and redshift $0.1-10$ km/s precisions. Although the latter could allow a spin constraint \citep{Kannan09,Angelil10b}, it is not achievable with current instruments, which are currently limited to $\sim 30$ km/s even for a star as bright as S2 \citep{Gillessen17}. 

Here we extend these studies by providing an expression for the detectability of a non-zero black hole spin as a function of stellar orbital parameters for a realistic GRAVITY observing campaign (duration, number of observations, achievable errors in astrometry and radial velocity). We use a semi-analytic geodesics code (\S\ref{sec:methods}) to rapidly simulate and fit relativistic orbits and show that it can reproduce past work on the star S2 (\S\ref{sec:code-validation-with}). We then simulate GRAVITY campaigns for closer in stars using astrometric data to determine the necessary conditions for a spin detection (\S\ref{sec:required-star-black}). From the current knowledge on the stellar distribution, we estimate the expected number of detectable stars satisfying these conditions (\S\ref{sec:expect-numb-stars}). We study improvements in the prospects of spin detection from combining astrometry with radial velocity measurements at a precision of $\sim 50$ km/s (\S\ref{sec:effect-radi-veloc}), more in line with what potentially could be reached with GRAVITY (Appendix \ref{app:A}). Discussion and conclusions are presented in \S\ref{sec:discussion}.

\section{Methods}
\label{sec:methods}

We approximate the potential near the black hole as a Kerr spacetime and stars as test particles. This is appropriate in the range $100 R_S \lesssim a \lesssim 5000 R_S$, where the lower and upper limits are set by the tidal disruption radius and Newtonian perturbations from the underlying stellar/remnant distribution, respectively \citep{Psaltis13,Merritt10,Zhang17}. We can then use geodesic ray tracing to follow the orbits of stars. We note that the upper limit could be much more constraining depending on the properties of the stellar/remnant distribution (see \S\ref{sec:discussion}).

\subsection{Stellar orbits as timelike geodesics}

We use the public \textsc{YNOGKM} code \citep{Yang14} to trace timelike geodesics in the Kerr metric. This code semi-analytically solves the geodesic equations in the Kerr metric by inverting the integral equation relating the $r$ and $\theta$ coordinates that results from separating the Hamilton-Jacobi equation \citep{carter1968}. The $\phi$ and $t$ coordinates are then given as elliptic integrals involving functions of $r$ and $\theta$ \citep{Rauch94}. The calculation of the many resulting elliptic integrals is sped up using the form developed by Carlson \citep{carlson1992,Dexter09}. The main difficulty with extending the method to timelike geodesics is accounting for the arbitrarily large number of $r$ turning points along the orbit. \citet{Yang13,Yang14} alleviate this problem by using a different independent variable, $p$, which monotonically increases from 0 to a maximum value along the geodesic.

To calculate the position of a star at coordinate time $t$ from an initial coordinate position and velocity (see below), we choose an initial guess of $p$ which is either equal to half of the maximum value along the geodesic (first point), or to the value used at the previous time (subsequent points). From the initial guess, a coordinate time is calculated, and the solution is then iterated until the observed time is found to the desired accuracy. Typically convergence to $10^{-6} GM/c^3$ is reached in $\lesssim 10$ iterations.

The \textsc{YNOGKM} code takes input initial position ($r_0$, $\theta_0$) in the coordinate frame, and the locally non-rotating frame \citep{Bardeen72} three-velocity, $v^{(i)}$. For comparison of our orbits with known and expected stars in the GC, it is most convenient to parameterize in terms of the Keplerian orbital elements ($a_{orb},e,i_{orb},\omega,\Omega,T_p$). We calculate approximate coordinate positions and velocities corresponding to the orbit by assuming that the star is non-relativistic near apocenter. The orbital elements are specified relative to the sky plane, while the input position and velocity to \textsc{YNOGKM} are relative to the black hole coordinate frame. We rotate the sky coordinates of the star to allow for arbitrary position angle and inclination of the black hole spin axis. With the convention that sky coordinates $(x,y,z)$ point along the RA, DEC and line of sight (away from the observer) directions, we define the black hole spin angles $i_{spin}$ ($[0,\pi]$) and $\epsilon_{spin}$ ($[0,2\pi]$) as the angle between the spin axis and $z$ and between the projection of the spin axis onto the sky plane and $-x$, respectively. 

This semi-analytic geodesic method is particularly efficient here. Each sample of a stellar orbit is independent, and so sampling at irregular, sparse observing epochs does not require integrating the orbit over many periods. This is the limit where analytic codes can be significantly faster than numerical integration while maintaining machine precision \citep{Dexter09}. 

\subsection{Redshift Calculation} 

The redshift of the received starlight is 

\begin{equation}
Z \equiv \frac{E_*-E_0}{E_0} = \frac{E_*}{E_0}-1 = \frac{\mathbfit{p}_* \cdot \mathbfit{u}_*}{\mathbfit{p}_0 \cdot \mathbfit{u}_0} -1
\end{equation}

\noindent where $E_*$ and $E_0$ are the photon energies measured by an observer co-moving with the star and at infinity, $\mathbfit{u}_*$ and $\mathbfit{p}_*$ are the star's four-velocity and the photon's four-momentum at photon emission, and $\mathbfit{u}_0 = (1,0,0,0)$ and $\mathbfit{p}_0$ are the four-velocity of the observer at infinity and the photon's four momentum at photon reception. 

The four-velocity $\mathbfit{u}_*$ is computed from the stellar orbits code, while $\mathbfit{p}_*$ can be computed from the impact parameters of the photon \citep{Cunningham73}. 

\subsection{The photon orbit} 

Light bending of photons affects both the measured position of the star as well as the redshift. Since the impact parameters of the photon are not known a priori, an exact calculation is costly and requires an iterative approach, in which e.g. the photon is propagated back from the observer until it passes close enough to the star \citep[e.g.][]{Zhang15}. We can simplify the problem considerably by noting that the effect of black hole spin on the photon orbit for the stars of interest in this paper ($a_{orb} \gtrsim 100 R_S$) is $\ll 1 \mu$as \citep{Bozza12,Zhang15} and $< 3$ km/s \citep{Angelil10b,Zhang15}, corresponding to $\lesssim 0.1\%$ and $\lesssim 10\%$ of the spin effects on the stellar orbit \citep{Zhang15}. Therefore, for the purposes of this paper, we can compute the photon orbit in the Schwarzschild metric. 

Furthermore, the weak-field approximation is valid for the orbits considered here, except for stars with extremely high inclinations as they pass behind the black hole. An upper limit to the closest approach distance of the photon to the black hole can be estimated as $d \sim R_p \cos(i_{orb})$, where $R_p = a_{orb} (1-e)$ is the periastron distance. For $R_p = 100 R_S$, the photon could pass closer than $20 R_S$ from the black hole for inclination $i_{orb} \goa 78 \degr$. For randomly oriented orbits, the probability of such high inclinations is very small at $1-\cos(90 \degr - 78 \degr) \sim 2 \%$. It is interesting to note, however, that if such a star is indeed found, light bending effects on astrometry and redshift during its passage behind the black hole could be quite significant and could potentially allow probing the black hole spin \citep{Bozza12}. 

For the Schwarzschild metric, it is possible to obtain a second-order differential equation relating $r$ and $\phi$ (photon coordinates in star-black hole-observer plane) which is independent of the photon's impact parameter: 

\begin{equation}
\frac{d^2u}{d\phi^2} + u = 3 G M u^2
\end{equation}

\noindent where $u=\frac{1}{r}$ \citep{GenRelWorkbook}. In the weak-field limit, an analytic perturbative solution can be derived: 

\begin{equation}
u(\phi) = A\sin(\phi+\phi_0) + \frac{3GMA^2}{2}+\frac{GMA^2}{2}\cos(2(\phi+\phi_0)) 
\end{equation}

\noindent where $A$ and $\phi_0$ are integration constants. Given the initial $(r_*,\phi_*)$ and final $(r_0,\phi_0) = (\infty,0)$ positions of the photon, this nonlinear equation can be solved numerically for $(A,\phi_0)$ and the impact parameter determined from 

\begin{equation}
\lim_{r->\infty} r \sin \phi = \frac{1}{A\cos(\phi_0)-GMA^2\sin(2\phi_0)},
\end{equation}

\noindent The impact parameters on the sky plane then give the measured astrometric position and redshift. 

\section{Code Validation with the Star S2} 
\label{sec:code-validation-with}

Of the currently known S-stars, S2 is the one with the closest approach to the black hole ($R_p \approx 1300 R_S$), and the potential to detect relativistic effects through the monitoring of its orbit has been the subject of numerous works as mentioned in Section \ref{sec:intro}. It therefore offers an opportunity to validate our code by comparing the measured relativistic effects with results from previous work. In all of the following, we adopt the mass of the black hole $M_{BH} = 4.3 \times 10^6 M_{\odot}$, the distance to the Galactic Center $R_0 = 8.3$ kpc and the following orbital parameters for S2: semi-major axis $a_{orb} = 111.1$mas, eccentricity $e=0.881$, inclination $i_{orb} = 131.9 \degr$, argument of periastron $\omega = 65.4 \degr$, longitude of ascending node $\Omega = 225.0 \degr$ and time of periastron passage $T_p = 2002.33$ years \citep{Gillessen09,Gillessen17}. We will also consider a hypothetical star with the same orbital parameters as S2 but a ten times smaller semi-major axis ("S2/10"). 

\subsection{Low-order Relativistic Effects} 

We checked that the orbit and redshift curves we obtain for S2 match the observed ones \citep{Gillessen17}. For a zero spin orbit, we checked that the periastron shift for S2 and S2/10 match the expected values $\delta \omega \Big|_{orbit} \approx 0.22 \degr$ and $2.2 \degr$ respectively. 

\begin{figure}
\includegraphics[width=\columnwidth]{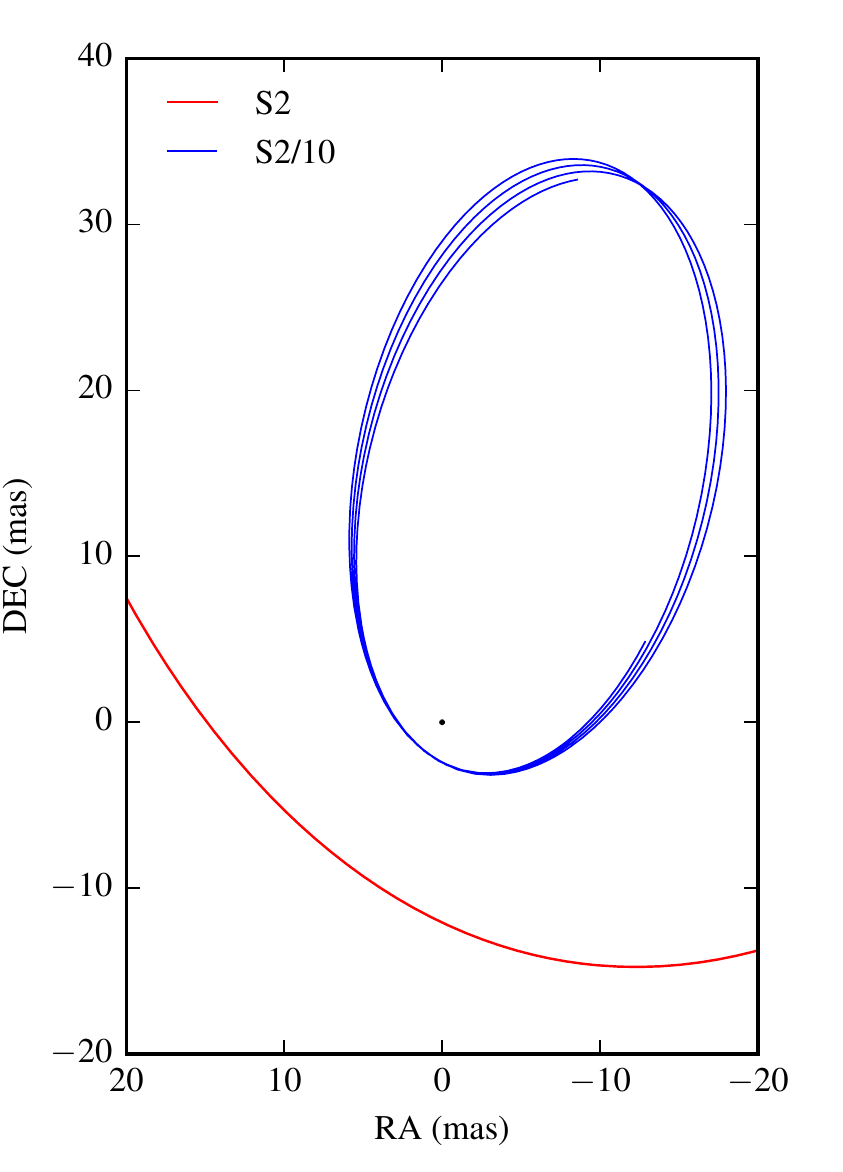} \\
\caption{Example of Schwarzschild precession for the hypothetical star "S2/10" during 5 years. A portion of the S2 orbit is also shown.}
 \label{fig:S2/10}
\end{figure}

\begin{figure*}
\includegraphics[width=2.2\columnwidth]{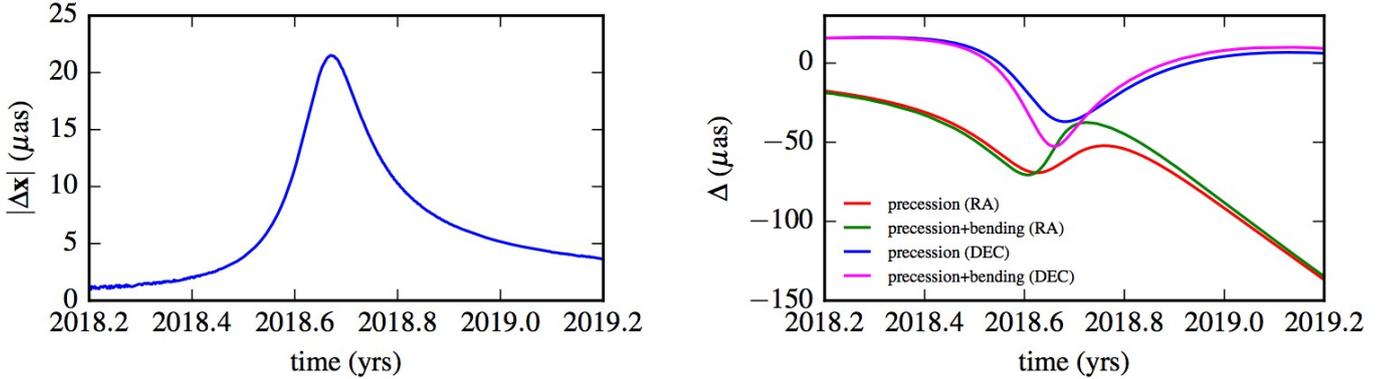} \\
\caption{\textbf{Left}: The effect of light bending during periastron passage of S2 peaks at $\approx 20\mu$as 12 days after periastron. This is the difference between an orbit with and without light bending. \textbf{Right}: Effect of periastron shift alone and with light bending around periastron passage of S2. Light bending enhances the periastron "kink" and could help in the early detection of the combined effect, before having to wait for the continuous growth in the years following periastron passage. In both cases, the curves are the difference between the relativistic (without or with light bending) and the Keplerian orbits.}
 \label{fig:bending}
\end{figure*}

Fig. \ref{fig:S2/10} shows an example of successive periastron shifts for S2/10 over a period of five years. Fig. \ref{fig:bending} shows the effect of periastron shift compared to a purely Keplerian orbit for S2 as a function of time near periastron passage, with the characteristic "kink" during periastron followed by the continuous increase of the effect over the following years. These curves are simply the difference between the relativistic and the Keplerian orbits for the same initial parameters. Similarly, we also tested the effects of transverse Doppler shift and gravitational redshift on the orbit of S2 during periastron, which amount to a maximum deviation of $\approx 100$ km/s each. These effects are all consistent with previous work \citep[e.g.][]{Weinberg05,Zucker06,Angelil10b,Grould17}.

\subsection{Photon Orbit} 

In order to test the implementation of our solution for the photon orbit, we computed the effect of light bending on the astrometric position of the star S2. This is done by computing the difference $|\sqrt{\text{RA}^2+\text{DEC}^2}|$ as a function of time between bent and non-bent photon orbits. As shown in Figure \ref{fig:bending}, the effect amounts to a deviation of $\approx 20\mu$as during periastron passage. This is consistent with previous results \citep{Bozza12,Grould17}. In Fig. \ref{fig:bending}, we also show the superimposed effects of periastron shift and light bending; the latter amplifies the former during the periastron "kink", enhancing the chances of an early detection of the combined effect. 

\subsection{Spin Effect} 

As mentioned above, for S2 the time scale associated with Newtonian perturbations due to an underlying stellar/remnant distribution is still shorter than the one associated with Lense-Thirring precession \citep{Merritt10,Zhang17}. Nevertheless, several works have estimated the astrometric effect of frame-dragging on the orbit of S2 \citep[e.g.][]{Zhang15,Yu16,Grould17}. In particular, \cite{Yu16} computed the shift in the apocenter position of S2 after one full orbit as a function of the two spin angles, as the effect can vary by more than an order of magnitude depending on the latter. We carried out a similar simulation by computing the maximum position difference $|\Delta \mathbfit{x}|$ between an orbit of S2 with $\chi=0$ and one with $\chi=0.99$ after one full period, where $\chi \in (0,1)$ is the normalized black hole spin. The resulting angle dependence and position difference ($1-15 \mu$as) are in agreement with \cite{Yu16}. Finally, we note that the deviation averaged over the spin angles, defined as 

\begin{equation}
\frac{1}{4 \pi} \int_0^{2\pi} \int_0^{\pi} |\Delta \mathbfit{x}| (i_{spin},\epsilon_{spin}) \sin{i_{spin}} d i_{spin} d \epsilon_{spin} 
\end{equation} 

\noindent is $\approx 8.5 \mu$as. For S2/10, the maximum astrometric shift over one full orbit as computed with the code is $\approx 3-48 \mu$as, with average over spin angles $\approx 27.5 \mu$as. 

We also computed the equivalent spin effects on the redshift. Again, the spin-angle dependence and size of the effects ($0.01-0.3$ km/s and $3-30$ km/s per orbit for S2 and S2/10, respectively) are consistent with \cite{Yu16}. 

\section{Required Orbital Parameters for a Black Hole Spin Measurement}
\label{sec:required-star-black}

Using closer stars to measure black hole spin overcomes the dominance of Newtonian perturbations over the frame-dragging precession time scale. The significance $\sigma$ of a spin detection through astrometry for an observing campaign of length $T$ years, total number of observations $N_{obs}$ and astrometric precision $\sigma_{x}$ scales as 

\begin{equation}
\sigma \propto \frac{\delta x \Big|_{T}}{\sqrt{N_{obs}} \sigma_x}
\end{equation}

\noindent where $\delta x \Big|_{T}$ is the total astrometric shift due to Lense-Thirring precession. The Keplerian elements which experience nonzero average changes due to black hole spin over a full orbit are all angles ($\omega$, $\Omega$ and $i_{orb}$) and the changes scale as 

\begin{equation}
\delta \omega, \delta \Omega, \delta i_{orb} \Big|_{orbit} \propto \chi \left(\frac{1}{a_{orb}(1-e^2)}\right)^{3/2} 
\end{equation} 

\noindent \citep[for full expressions, including dependence on spin angles, see for e.g.][]{Iorio11}. The \textit{astrometric} change scales with the size of the orbit $a_{orb}$

\begin{equation}
\delta x \Big|_{orbit} \propto \chi \frac{1}{a_{orb}^{1/2}(1-e^2)^{3/2}}
\end{equation} 

\noindent The effect per orbit is therefore not strongly dependent on $a_{orb}$. The main benefit of using closer stars is their shorter periods, which, combined with the fact that precession is a cumulative effect, lead to a more significant astrometric deviation over a fixed period of time. Since the orbital period $P_{orb} \propto a_{orb}^{3/2}$, the astrometric deviation over a fixed time $T$ scales as  

\begin{equation}
\delta x \Big|_{T} \propto \frac{T}{P_{orb}} \chi \frac{1}{a_{orb}^{1/2}(1-e^2)^{3/2}} \propto \chi T \frac{1}{a_{orb}^{2}(1-e^2)^{3/2}}
\end{equation}

\noindent We therefore have 

\begin{equation}
\label{eq:scaling}
\sigma \propto \frac{\chi T}{\sqrt{N_{obs}}\sigma_x}{a_{orb}^{2}(1-e^2)^{3/2}} 
\end{equation}

\noindent A similar expression is used in \cite{Weinberg05} but parametrized in terms of number of orbits covered instead of observing campaign duration. Our goal is to determine the properties of a star $(a_{orb},e)$ that would allow a measurement of black hole spin with a realistic campaign with duration $\lesssim$ 10 years, and subsequently infer the expected number of stars satisfying such constraints based on what is currently known about the stellar distribution in the innermost arcseconds of the Galaxy. We therefore expect $\sigma$ contours to have the form 

\begin{equation}
\label{eq:contour} 
a_{orb} (1-e^2)^{3/4} = R 
\end{equation} 

\noindent where $R$ depends not only on the spin and observing campaign parameters but also on effects such as the masking of spin-related effects by fitting of the remaining parameters. We use simulated stellar orbits in order to estimate such unknown normalization factors. 

\subsection{Simulated Stellar Orbits} 

The astrometric deviations due to spin cited above were calculated as the difference between models with zero and maximum black hole spin when keeping all other parameters (initial positions and velocities of the star, BH mass and distance) constant. In practice, such parameters are not exactly known and have to be fit together with the black hole spin, which leads to masking of the spin-related effects. 

We simulate stellar orbits across a grid of $(a_{orb},e)$, with $a_{orb} \in (200 R_S, 5000 R_S)$ and $e \in (0.1,0.9)$. The other Keplerian parameters ($\omega$, $\Omega$, $i_{orb}$) are taken to be the same as for S2. They only matter in relation to the black hole spin angles as far as spin-related effects are concerned. We choose $i_{spin} = \epsilon_{spin} = 0 \degr$, which give astrometric shifts close to the average over the spin angles (specifically, $9.6$ and $31.7 \mu$as/orbit for S2 and S2/10, respectively, compared to the averages of $8.5$ and $27.5 \mu$as referred above). We use $\chi = 0.9$ and the canonical $M_{BH}$ and $R_0$ as above. The observing campaign is set to a total duration of $T=4$ years, with observations taken in $3$ consecutive months per year over a period of $10$ consecutive days per month, for a total of $N_{orb}=120$ observations. Gaussian errors with $\sigma_x =10 \mu$as and $100 \mu$as are added representing canonical astrometric accuracies that could be achieved with GRAVITY. 

Given the simulated measurements with errors, one approach that could be used to estimate the spin error is to perform a Bayesian fit to the data using e.g. Markov chain Monte Carlo approaches \citep{Zhang15,Yu16}. However, we have found that ensuring the convergence of the spin parameters with such an approach is very challenging as the orbit is much more sensitive to other non-spin parameters, which must be simultaneously fit and which can also mask the effects of spin. We instead use a null hypothesis test in order to estimate the significance of spin, which entails in finding the best-fit zero-spin orbit to the simulated data and assessing the probability that the data were generated by such a model based on the residuals. For this, we follow the usual approach of converting the $\chi^2$ of the best-fit solution to a p-value, and that into a significance $\sigma$:

\begin{equation}
\sigma = \sqrt{2} \erfinv(1-$p-value$)
\end{equation} 

In order to find the best-fit zero-spin solution, we used a downhill simplex algorithm \citep[][]{Nelder65} which does not require numerical derivatives (as opposed to gradient methods such as Levenberg-Marquardt) and was found to be more stable and less sensitive to local minima. The method is based on constructing simplexes (polytopes of $n+1$ vertices in $n$ dimensions) and updating the vertices with operations (reflection, expansion, contraction, shrinkage) which result in successively better solutions. Because the method is not completely immune to local minima, we used $10$ initial simplexes distributed around the initial parameters for each fit. This number was found to be sufficient in order to avoid local minima. We also note that no priors were used for $M_{BH}$ and $R_0$. Even though they are constrained by the currently known S-stars, the best fit zero-spin solution is always within a few percent of the initial value and therefore the currently known bounds would not lead to a change in significance. 

\subsection{Results} 

\begin{figure*}
\includegraphics[width=2.2\columnwidth]{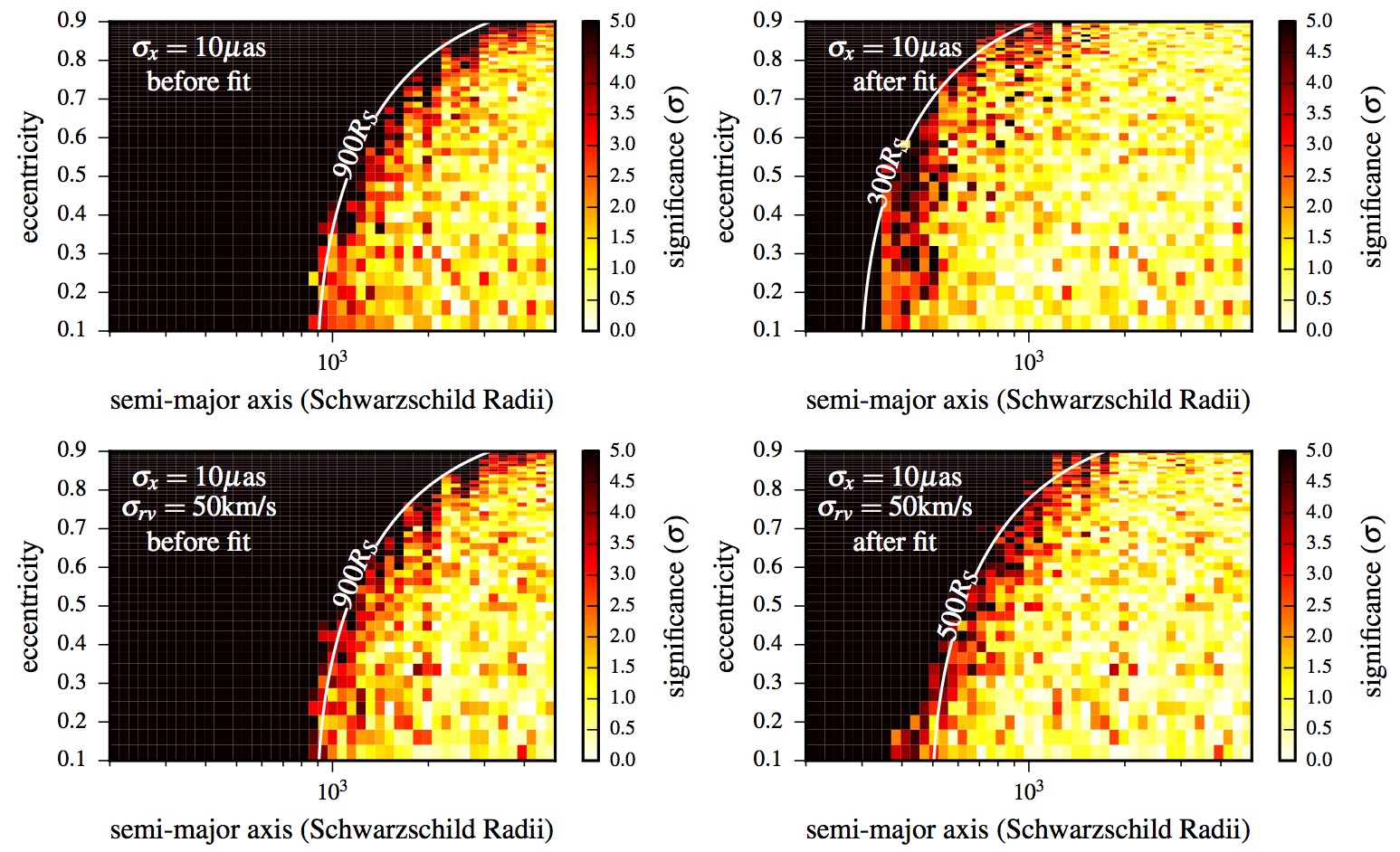} \\
\caption{Contour plots for the significance $\sigma$ of a black hole spin detection through monitoring of stellar orbits as a function of $(a_{orb},e)$ for an observing campaign of 4 years with a total of $N=120$ observations. The black hole normalized spin is $\chi=0.9$ and spin angle parameters that lead to an average astrometric deviation are assumed. The white lines show the expected contour for astrometric deviations due to spin effects, $a_{orb}(1-e^2)^{3/4} = \text{ constant}$, which separates the regions of low and high significance. The left panels show $\sigma$ before finding the best fit zero-spin solution i.e. simply setting $\chi=0$, whereas the right panels show $\sigma$ after fitting for the best zero-spin solution. The upper panels are for a purely astrometric campaign with precision $\sigma_x = 10\mu$as, while the lower panels contain additional radial velocity measurements with precision $\sigma_{rv} = 50$ km/s. Having to fit all parameters leads to masking of the spin-related relativistic effects and leads to more stringent limits on the required star for a spin detection. The additional radial velocity measurements do not lead to an increase in significance before fitting, but help with constraining the non-spin parameters during the fit, ameliorating their masking of the spin effects.}
 \label{fig:contours}
\end{figure*}

Fig. \ref{fig:contours} (top row) shows the resulting significance contours $\sigma (a_{orb},e)$ for $\sigma_x = 10 \mu$as. The graininess arises due to different error instantiations between simulated orbits. There is a region of low spin significance ($\sigma \lesssim 2$) and of high spin significance ($\sigma \gtrsim 5$) separated by a relatively narrow transition region. The white contour lines show analytic estimates for the shape of this region based on Eq. (\ref{eq:contour}). 

The left panel shows the significance before fitting i.e. using the initial parameters but setting $\chi=0$, whereas the right panel shows the significance after finding the best fit zero-spin solution. The pure size of the effect would suggest a star with $a_{orb} (1-e^2)^{3/4} \lesssim 900 R_S$ is needed to detect spin at high significance, but in practice when fitting for all parameters a star with $a_{orb} (1-e^2)^{3/4} \lesssim 300 R_S$ (i.e. $\sim 3 \times$ closer in) is required. 

Following Eq.(\ref{eq:scaling}), we can then write the requirement on $(a_{orb},e)$ of the star as a function of the observing campaign length $T$ in years, total number of observations $N_{obs}$, astrometric error $\sigma_x$ in $\mu$as and normalized spin magnitude $\chi$ as 

\begin{equation}
a_{orb}(1-e^2)^{3/4} \lesssim 300 R_S \sqrt{\frac{T}{4 \text{years}}} \left(\frac{N_{obs}}{120}\right)^{0.25} \sqrt{\frac{10 \mu as}{\sigma_x}} \sqrt{\frac{\chi}{0.9}}
\end{equation}

We have repeated the same experiment using $\sigma_{x} = 100 \mu$as. In accordance with the expression above, we found $a_{orb} (1-e^2)^{3/4} \lesssim 100 R_S$ is needed for a high significance spin detection.  

\subsection{Expected Number of Stars} 
\label{sec:expect-numb-stars}

In order to translate the constraint on orbital parameters from the previous section into an expected number of stars for which GRAVITY would be able to detect black hole spin, it is necessary to estimate the probability densities of semi-major axis, $n(a_{orb})$, and eccentricity, $n(e)$, and the K-band Luminosity Function (KLF, $\frac{d\log N(K)}{dK} = \beta$) in the central 1"/0.04pc. 

The latter has been estimated by several works with the consistent result $\beta \approx 0.20$ \citep{Genzel03, Buchholz09, Sabha12}. The most recent analysis of the S-stars orbits is consistent with a "thermal" eccentricity distribution \citep{Gillessen17}; we therefore adopt $n(e) de = 2 e de$. For such a distribution, if the energy distribution function follows a power-law, $f(\epsilon) \propto \epsilon^p$, then the space density distribution $n(r) \propto r^{-\gamma}$ and $n(a_{orb}) \propto a_{orb}^{-\gamma+2}$ \citep{Schoedel03}. Estimates of $\gamma$ from stellar counts in the region $r \lesssim 10"$ consistently find $\gamma \approx 1.2-1.4$ \citep{Genzel03, Schoedel07, Do09}.  We therefore adopt $n(a_{orb}) da_{orb} \propto a_{orb}^{0.7} da_{orb}$, which is also in accord with the estimated semi-major axis distribution directly from the orbits of S-stars \citep{Gillessen09}. 

We generate a mock field of $N=10^7$ stars with Keplerian orbits with $a$ and $e$ following the distributions above, using inverse transform sampling on the respective Cumulative Distribution Functions (CDFs): 

\begin{flalign}
   &N_E(e) = e^2 & \\
   &N_A(a_{orb}) = \left(\frac{a_{orb}}{a_{max}}\right)^{1.7}
\end{flalign}
\\
$a_{max}$ is set to 10" ($n(a_{orb}) =0$ for $a_{orb}>a_{max}$), which is justified below. The time of periastron is set randomly between $0$ and $10^6$ years, which is larger than the largest possible period ($\approx 12,000$ years). Consistently with the current known S-stars, the orbital orientations are drawn isotropically i.e. $\omega$ and $\Omega$ have uniform distributions and $n(i_{orb})di_{orb} = \frac{1}{2} \cos(i_{orb}) di_{orb}$. We exclude stars that would undergo tidal disruption i.e. for which $R_p = a_{orb}(1-e) < R_t = R_* \left(\frac{M_{BH}}{M_*}\right)^{1/3} \sim 30 R_S$. The effective number of stars is then normalized by the number of stars within a radius $r<1"$, which is known to be $\approx 56$ for $K\lesssim17$ \citep{Genzel03}. The final expected number for a given magnitude limit is then estimated using the KLF. 

As mentioned, we set $a_{max} = 10"$ since $n(a_{orb})$ would otherwise diverge. There is evidence from stellar counts that $\gamma \sim 2$ for $r>10"$ \citep{Genzel03, Schoedel07,Fritz16}, so that $n(a_{orb}) = \text{constant}$ for $r>10"$. Furthermore, beyond the radius of influence of the black hole $\sim$75" \citep[][we also note the direct measurements of a half-light radius of the nuclear star cluster $\sim 100"/178"$ by \cite{Schoedel14} and \cite{Fritz16}]{Alexander05}, the orbits are significantly perturbed and not dominated by the gravitational potential of the black hole anymore. To check whether the chosen $a_{max}$ leads to a bias, we simulate the contribution from stars with $10"<a_{orb}<75"$ to the region $r<1"$, which is found to be $\lesssim 1\%$, since stars that do have the potential to reach $r<1"$ due to their higher eccentricities spend a very small portion of their orbital periods in this region. 

We choose $K<19$ as the upper limit on a star which could be detected with GRAVITY \citep{Eisenhauer11}. The simulation predicts $\sim 1$ such star within a radius $r<50$ mas ($\sim$ FOV of GRAVITY), in agreement with previous estimates \citep{Genzel03}. The median $(a_{orb},e)$ of such a star from the simulations is $\approx$(80 mas, 0.8). Because we have included the eccentricity distribution, we can also predict the expected number of stars that satisfy the contours for measuring spin derived in the previous section. We include an additional cut-off in $a_{orb}$ in order to ensure that at least one orbit is covered by the observing campaign ($a_{orb}<5000 R_S$ for $T=4$ years). 

Table \ref{tab:number} shows the expected number of stars which would allow a detection of spin. The masking of the relativistic effects by other parameters ("Fit" vs "No Fit") leads to a significant reduction in the expected number of stars. For our canonical observing campaign of $T=4$ years, $N_{obs}=30\times4=120$ and $\sigma_x = 10\mu$as, we expect $0.035$ star that would allow a significant detection of black hole spin. The median $(a_{orb},e)$ of such stars is $\approx(1200 R_S, 0.95)$. 

We note that more recent papers have studied the faint population of stars in the Galactic Center in more detail, 
both using stellar counts going down to fainter limits \citep{Gallego18} as well as the faint diffuse light \citep{Schoedel18}. It is important to 
compare the estimated numbers above (which used assumptions on the stellar population based on brighter stars) to the ones based on the faint 
population alone. They found $\sigma_0 = 20 \text{ stars } \text{arcsec}^{-2}$ and $\sigma_0 \sim 72 \text{ stars } \text{arcsec}^{-2}$ at $R_0 = 0.25"$ 
for stars with $17.5 \leq K \leq 18.5$ and $18.5 \leq K \leq 19.5$, respectively, and a surface density exponent $\Gamma \sim -0.4$. From that we 
estimate $\approx 0.5$ and $\approx 2$ stars within 50 mas in the two magnitude ranges above, compared to the $\approx 1$ star with $K<19$ we found before. 
In this case, it is possible that our numbers above are pessimistic to within a factor of $\sim 2$. We also note that \cite{Schoedel18} and \cite{Gallego18}
conclude that the population of faint stars is likely to be dominated by the old star population, and in that case a possible faint star found by 
GRAVITY is more likely to be part of the old cusp rather than a faint, typically young S-star of type A.

One of the consequences of Eq.(\ref{eq:scaling}) is that increasing $T$ and $N_{obs}$ or decreasing $\sigma_x$ does not strongly increase the limit on $a_{orb}(1-e^2)^{3/4}$ (and therefore the number of stars). If we instead consider an observing campaign of $10$ years with $N_{obs} = 30 \times 10=300$ total observations, then using Eq.(\ref{eq:scaling}) (and requiring $a_{orb}<9000 R_S$ so that at least one orbital period is covered), the expected number of stars for measuring black hole spin increases to $0.12$, with the median $(a_{orb},e)$ of such stars $\approx (2400 R_S,0.96)$. The fraction of such stars with $a_{orb}>5000 R_S$ (for which a spin measurement would very likely start to suffer from Newtonian perturbations) is $\approx 25\%$. 

\begin{table}
 \caption{Expected number of stars with $K<19$ within given contour regions for measuring black hole spin for an observing campaign of duration $T$ with 30 observations/year, $\chi=0.9$ and spin angle parameters that lead to an average astrometric deviation.}
 \label{tab:number}
 \begin{tabular}{|c|c|c|c|}
 \hline
 & \shortstack{No Fit \\ T=4 years} & \shortstack{Fit \\ T=4 years} & \shortstack{Fit \\ T=10 years} \\ \hline
 \shortstack{$\sigma_x = 10 \mu$as \\ no rv} & 0.15 & 0.035 & 0.12 \\ \hline
 \shortstack{$\sigma_x=10 \mu$as \\ $\sigma_{rv} = 50$km/s} & 0.15 & 0.07 & 0.23 \\ \hline
\end{tabular}
\end{table}

We can also use these scalings to estimate the potential of future extremely large telescopes for measuring black hole spin from stellar orbits. The main advantage is the large increase in sensitivity. Assuming a limit $K<22$, the numbers above should be multiplied by a factor of $4$ based on the KLF. However, if the astrometric precision cannot reach the $10\mu$as level, the number of stars would be reduced accordingly. Also, if such telescopes could reach radial velocity precisions $\sim 1-10$ km/s on faint stars, those could also be used to probe the black hole spin. Because the radial velocity changes are strongest near periastron (as opposed to apastron for the astrometric changes), they should be more robust to Newtonian perturbations \citep[e.g.][]{Psaltis16}. Finally, in the case of a large FOV there is the possibility of measuring relativistic effects from the collective motions of many further out stars \citep[e.g.][]{Do17}, but such an approach may be a challenge for measuring spin due to Newtonian perturbations. 

We note that these estimates should be taken with caution since they are based on an extrapolation to the very inner region around the black hole which has been beyond the resolution limits of any instrument before GRAVITY. A cusp of massive stars in the immediate vicinity of the black hole \citep{Alexander09}, precursors of stellar-mass black holes, for example could increase the expected number. Alternatively, a break in the KLF from bursts of star formation history \citep{Pfuhl11} could decrease it. 

\section{Effect of Radial Velocities}
\label{sec:effect-radi-veloc}

In the above analysis for detecting black hole spin, we have not so far considered radial velocity measurements, which 
could potentially be made by measuring the redshifts of spectral lines as is traditionally done for S-stars in the 
Galactic Center \citep{Gillessen09}. In order to probe redshift effects on a stellar orbit due to spin, however, 
a very high redshift precision would be needed. As mentioned above, the maximum redshift difference (assuming the 
most optimistic spin angles) over a full orbit between models with $\chi=0$ and $\chi=0.99$ is $\approx 0.3$ km/s 
and $30$ km/s for S2 and S2/10, respectively. The effect is also extremely 
sharped around periastron passage, and would require very targeted observing campaigns in order to be detected.

Previous work \citep{Zhang15,Yu16} assumed redshift errors $1-10$ km/s combined with astrometry when estimating spin errors. 
Such precision allows redshift probes of spin for potential stars in close orbits around SgrA*, but could only be achieved 
with future facilities such as E-ELT or TMT \citep{Do17}. Current redshift measurements for the star S2 \citep[significantly brighter than any potential close star,][]{GRAVITY17} have uncertainties $\sim 30$ km/s. Therefore, a natural question to ask is the extent to which potential radial velocity measurements 
with errors $\sigma_{v} > 30 $ km/s could help in detecting spin. Although such precision would likely not be enough to measure spin by itself, it should 
help in better constraining the other parameters and therefore prevent the masking of spin effects to some extent, alleviating the constraints on the required stellar orbits. 

\begin{figure*}
\includegraphics[width=2.2\columnwidth]{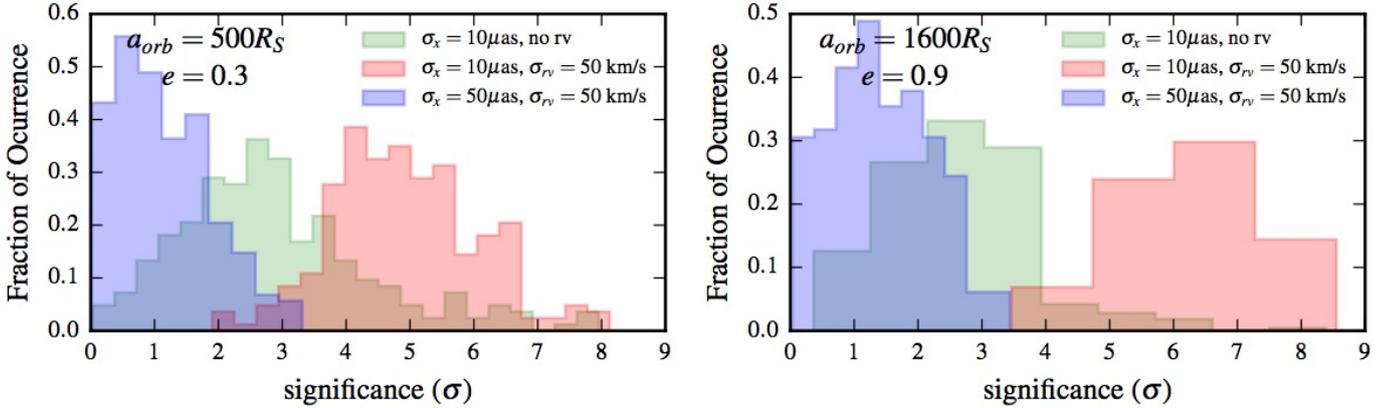} \\
\caption{Effect of radial velocity measurements on the significance $\sigma$ of black hole spin. We used the canonical observing campaign discussed previously, selected two orbits near the contour line ($a(1-e^2)^{3/4} \approx 460 R_S$) of the upper right plot of Figure \ref{fig:contours}, and ran $240$ simulations to determine the significance after finding the best fit zero-spin solution. This was repeated for three cases: only astrometry with $\sigma_x = 10\mu$as (i), and including radial velocity measurements with $\sigma_{rv}=50$ km/s with (ii) and without (iii) degradation of the astrometric precision to $\sigma_{x}=50\mu$as. The spread in $\sigma$ for each case is due to the different error instantiations. Adding radial velocity measurements increases the significance by better constraining the orbital parameters, as long as the astrometric precision can be maintained. Notice also that the increase in $\sigma$ due to radial velocities is more pronounced for the eccentric orbit.}
 \label{fig:histograms}
\end{figure*}

Besides direct spectroscopy, we suggest a potential method to measure radial velocities of close orbit stars directly with GRAVITY using spectral 
differential interferometry in medium resolution. The method is outlined in Appendix \ref{app:A} and we estimate a radial velocity precision 
$\sigma_v \sim 50$ km/s. In order to test the effect of adding the radial velocity measurements, we selected two example orbits near the transition region 
in Fig. \ref{fig:contours}, $(a_{orb},e) = (500 R_S, 0.3)$ and $(1600 R_S, 0.9)$, corresponding to $a(1-e^2)^{3/4} \approx 460 R_S$, and ran a series of $240$ fits using (i) only astrometric measurements with $\sigma_x = 10 \mu$as; (ii) astrometric and radial velocity measurements with $\sigma_x = 10 \mu$as and $\sigma_v = 50$ km/s; and (iii) same as previous but with $\sigma_x = 50 \mu$as. The latter covers the case of radial velocity measurements at the expense of astrometric precision. Figure \ref{fig:histograms} shows the histograms for the significance $\sigma$ of spin detection for each case. The spread in $\sigma$ (related to the graininess of the contour plots shown before) is natural due to the different error instantiations. A clear trend is observed, in the sense that adding the radial velocities increases the significance of a spin detection only if the astrometric precision can be maintained. Otherwise, if the astrometric precision is degraded by a factor of a few, the significance is severely lessened as the radial velocities themselves do not probe spin effects, but rather lead to a better constraint on the other parameters. Moreover, the increase in significance when adding radial velocities is stronger for the more eccentric orbit. 

In order to estimate how the required number of stars changes when including radial velocities, we created $(a_{orb},e)$ contour plots with the same parameters as before but with additional redshift measurement $\sigma_{rv} = 50$ km/s at each observation. The result is shown in Fig. \ref{fig:contours} (lower row) and Table \ref{tab:number}. Whereas the contour line does not change before fitting for the zero-spin solution (since radial velocities with this precision are not probing spin effects), it does move outwards after performing the zero-spin solution fit, consistent with the behavior observed in the two specific examples above. For the canonical observing campaign we use, this means $a_{orb} (1-e^2)^{3/4} \lesssim 500 R_S$, with the expected number of stars increasing by a factor of two to $0.07$. Increasing the observing campaign to $T=10$ years increases the number of stars to $0.23$. 

\section{Discussion and Conclusion}
\label{sec:discussion}

In this paper, the main goal was to determine the requirements that a hypothetical star would need to satisfy in order to allow a detection of black hole spin through 
the astrometric monitoring of its orbit around the supermassive black hole in the Galactic Center. In order to do this, we made use of a semi-analytical Kerr geodesics code to calculate stellar orbits close to Sgr A*. Given the sparse sampling of the orbits and the large number of simulations needed, avoiding the numerical integration of the orbits leads to a significant improvement in computational time. For the photon orbit, because the spin effects on both astrometry and redshift are negligible compared to realistic precisions, we have used a weak-field Schwarzschild approximation. 

We tested the validity of our code by checking it reproduces the expected relativistic effects on the star S2 and a hypothetical star S2/10. In particular, we noticed that the light bending when S2 passes behind the black hole during periastron passage enhances the signature from periastron advance and could lead to an early detection of the combined effect. 

In order to detect black hole spin, closer stars are needed to overcome gravitational perturbations from other stars/remnants \citep{Merritt10,Zhang17}. For average black hole spin angle parameters, we found that the hypothetical star would have to satisfy 

\begin{equation}
a_{orb}(1-e^2)^{3/4} \lesssim 300 R_S \sqrt{\frac{T}{4 \text{years}}} \left(\frac{N_{obs}}{120}\right)^{0.25} \sqrt{\frac{10 \mu as}{\sigma_x}} \sqrt{\frac{\chi}{0.9}}
\end{equation}

\noindent in order to allow detection of black hole spin. The left hand side was derived analytically (Eq. \ref{eq:contour}). By fitting simulated relativistic orbits with errors and determining the significance of the deviation from the best-fit zero-spin solution, we verified this expectation and calculated the normalization of the right hand side. In particular, we found that having to fit for all parameters simultaneously leads to a requirement which is about a factor of three more stringent than what would have been predicted directly from the size of the relativistic effect. Using the current distributions of eccentricities and semi-major axis in the inner region of the GC, we have estimated that the number of stars satisfying this conditions is $\sim 0.035$ and $\sim 0.12$ for a 4-year and 10-year observing campaign, respectively, with 30 observations/year and $10 \mu$as errors in both cases. 

We have also shown the effect that radial velocities with precision $\sim 50$ km/s would have in the detection of spin. Although such redshift precision does not allow to probe spin directly, it helps by providing stronger constraints on the other parameters. The number of expected stars increases by a factor of two if radial velocities at this precision are also available. It is therefore important to consider the possibility of radial velocity measurements in parallel to astrometry, and we give an example of a potential method to do this with GRAVITY. 

In the above analysis, we have assumed that the black hole always lies at the origin. In practice, fitting S-star orbits requires the inclusion of offset and linear drift parameters of the black hole \citep{Boehle16,Gillessen17}. In the case of GRAVITY, astrometric measurements are taken relative to a reference star that is used to fringe track \citep{Eisenhauer11}. Detecting relativistic effects would then require measuring its orbital parameters. However, \citet{GRAVITY17} showed that it may be possible to reference stellar positions directly to Sgr A*, removing the need for reference frame parameters as long as the near-infrared emission originates close to the black hole. In the case of upcoming extremely large telescopes, the results obtained here still apply except that the reference frame parameters should be included.

We have assumed a pure Kerr metric for the spacetime around Sgr A*, corresponding to the most optimistic case. 
In practice, a distribution of stars/remnants introduces perturbations that could mask the precession due to black hole spin. 
Even though these two effects could potentially be separated \citep[e.g. astrometric deviations due to spin are maximum at 
apastron whereas Newtonian perturbations peak during periastron,][]{Zhang17}, disentangling them with limited observations 
and without prior knowledge on the perturbers would be very challenging. 
From the diffuse light background from faint stars that cannot be currently resolved, \cite{Schoedel18} estimated a total enclosed 
stellar mass of $\sim 180 M_{\odot}$ within 250 mas. Using their measured 3D power-law density profile $\gamma \approx 1.1$, this 
translates to only $\sim 2 M_{\odot}$ within 25 mas. From Fig.1 of \cite{Merritt10}, this would put an upper limit on $a_{orb} \sim 10,000 R_S$
for frame dragging dominating over Newtonian perturbations, which the stellar orbits considered in this paper are well within . 
However, both theoretical considerations as well as simulations predict an accumulation of stellar-mass black holes ($\sim 10 M_{\odot}$) 
in a steep cusp ($\gamma \approx 1.75-2.0$) close to SgrA* through mass segregation \citep{Freitag06,Hopman06,Alexander09,Preto10,Amaro-Seoane11}. 
Current upper limits on the total mass within 25 mas \citep[$1.3 \times 10^5 M_{\odot}$,][]{Boehle16} or 13 mas \citep[$4 \times 10^4 M_{\odot}$,][]{Gillessen17}
from fitting of stellar orbits cannot exclude the presence of a more massive dark cusp, and therefore it is important to consider its potential disturbance for the measurement of black hole spin with a closer star.

Both \cite{Merritt10} and \cite{Zhang17} consider a variety of stellar/remnant distributions with different perturber masses, total masses and density profiles
to compare the effect of Newtonian perturbations to black hole precession. While \cite{Merritt10} used N-body simulations to 
study the overall evolution of an entire cluster, \cite{Zhang17} studied the detailed evolution of a test star in response to the perturbers. 
We can use their results to assess how much a black hole spin measurement for our two example stars in \S\ref{sec:effect-radi-veloc} with 
$(a_{orb},e) = (500 R_S, 0.3)$ and $(1600 R_S, 0.9)$ would 
suffer from Newtonian perturbations for different cluster properties. Considering stars with  $e=0.88$ and $e=0.3$ with $\chi=1$ and spin angles
such that the spin-induced astrometric changes are average (as we consider here), \cite{Zhang17} find that the critical semi-major axes are 
$1440 - 2040 R_S$ and $1200 - 1560 R_S$ for a cluster of $10 M_{\odot}$ black hole perturbers with density profile $\gamma = 1.75$ and 
total mass of $30$ and 100 $M_{\odot}$ within 25 mas, respectively. Therefore, the more eccentric star would start to suffer from Newtonian 
perturbations for cluster masses $\gtrsim 100 M_{\odot}$. Since \cite{Zhang17} found that the effect of Newtonian perturbations is much less 
dependent on eccentricity than frame dragging, the advantage of a more eccentric star in terms of allowing for a larger semi-major axis is balanced 
by a higher sensitivity to Newtonian perturbations. Alternatively, from Fig. 3 of \cite{Merritt10}, for a cusp of $10 M_{\odot}$ black holes with $\gamma=2$
and a total mass $\lesssim 100 M_{\odot}$ within 25 mas and $\chi=1$, the critical radius is $\gtrsim 5000 R_S$, assuming the most optimistic 
black hole spin angles. Although the exact value depends on the black hole spin parameters, these results show that a steep cusp of black holes with total mass 
$\gtrsim 100 M_{\odot}$ within 25 mas could start to compromise the measurement of black hole spin with a potential closer star.

Other methods could also be used to detect the black hole spin of SgrA*. Pulsar timing could reach much higher precision \citep[e.g.][]{liuetal2012,Psaltis16,Zhang17b}, but the lack of ordinary pulsar detections in deep surveys of the central parsec \citep{johnstonetal1995,macquartetal2010,whartonetal2012,Dexter14} poses a significant challenge to this approach. Direct imaging of emission surrounding the "black hole shadow'' of Sgr A* with radio VLBI \citep{Doeleman09} could also potentially constrain spin, but so far suffers from complicated model-dependence \citep[e.g.][]{Broderick09,Dexter09b}. Finally, depending on the mechanism behind the NIR flares of SgrA*, astrometric monitoring of e.g. an orbiting hot spot could also allow a measurement of spin \citep[e.g.][]{Broderick06,Hamaus09,Vincent11}. All these methods are complementary. Although each is challenging, their combination could probe the spacetime around Sgr A* on scales ranging from $\sim 1-3000$ Schwarzschild radii.

\section*{acknowledgements}
JD thanks \citet{Yang14} for making the \textsc{ynogkm} code public and D. Psaltis and J. Stone for useful discussions. This work was supported in part by a Sofja Kovalevskaja Award from the Alexander von Humboldt Foundation of Germany.
\footnotesize{
\bibliographystyle{mnras}
\bibliography{mybib}
}

\appendix 

\section{Measuring Radial Velocities of a Faint Star with GRAVITY} 
\label{app:A}

\indent

Here, we explore the possibility of measuring radial velocities of a faint star with GRAVITY using differential visibility signatures across spectral lines. This would require the use of medium resolution ($R\approx500$). Although low resolution ($R\approx22$) has been the envisioned mode of operation in the Galactic Center due to SNR considerations, a bright flare state ($K\sim15$) or partial coupling of S2 ($K\approx14$) into the GRAVITY fiber \citep{GRAVITY17} could provide the necessary SNR for medium resolution, together with long integration times. 

A possible faint star is expected to have absorption lines in its spectrum. The faintest early-type stars observed spectroscopically in the Galactic Center ($K\lesssim 17.5$) are compatible with a A0/B9V classification and contain a Br$\gamma$ absorption line \citep{Pfuhl11}. For a fast moving star with $v_r \sim 10,000$ km/s, such a line would be significantly displaced from its rest wavelength; therefore, discovering differential visibility signatures at unexpected wavelengths could allow to identify such stars and measure their radial velocity. We focus on an early-type star since they dominate the current spectroscopically identified stars in the S-star cluster \citep{Eisenhauer05,Habibi17}; however, we note that late-type giants are also a possibility for faint stars \citep{Pfuhl11}, and could very well dominate the population of faint stars close to the center \citep{Schoedel18,Gallego18}. In the latter case, the series of sharp CO bands could provide even more convincing differential visibility signatures. 

\begin{figure}
\includegraphics[width=\columnwidth]{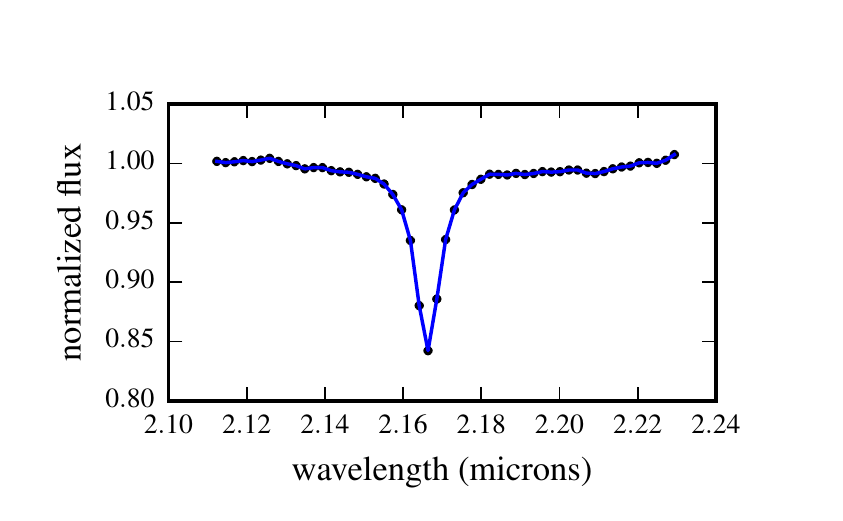} \\
 \includegraphics[width=\columnwidth]{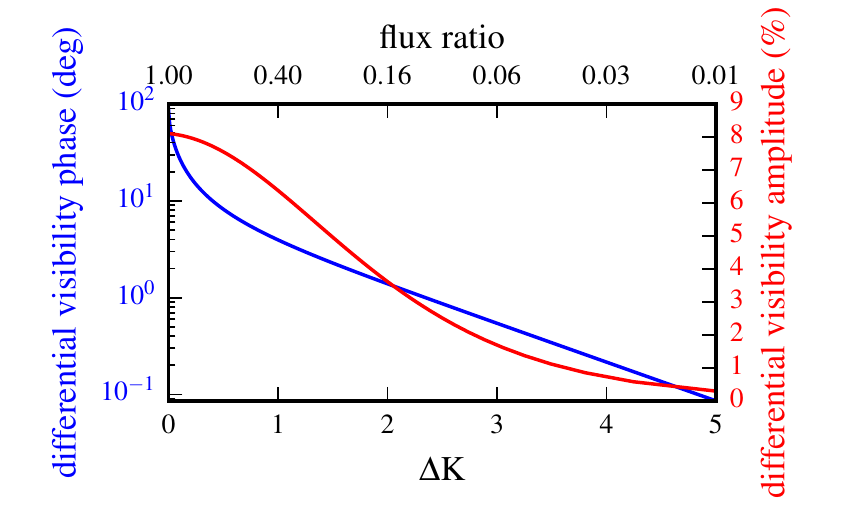}
 \caption{\textbf{Top:} Br$\gamma$ line of an A0V star \citep{Wallace97} with sampling appropriate to GRAVITY's medium resolution ($R=500$) mode. \textbf{Bottom:} Maximum differential visibility phase and amplitude across a spectral line for a binary as a function of the flux ratio or near infrared K band magnitude difference. We assume a line depth $0.85$ of the continuum corresponding to the spectrum above.}
 \label{fig:differential_size}
\end{figure}

In order to simulate the expected size of the differential visibility signatures and the precision of the resultant radial velocity measurement, we consider a binary system consisting of a brighter source (either a flare or the star S2 displaced from the center of the GRAVITY fiber PSF) and a fainter source corresponding to the fast-moving star. The complex visibility of a binary system is 
\begin{align}
V(\bmath{u}\cdot \bmath{\sigma},f) = \frac{1+f e^{-2\pi i \bmath{u} \cdot \bmath{\sigma} }}{1+f}
\end{align} 
where $f$ is the flux ratio, $\bmath{\sigma}$ the separation vector and $\bmath{u}=\frac{\bmath{B}}{\lambda}$ the spatial frequency, and has unit period in $\bmath{u}\cdot\bmath{\sigma}$. Figure \ref{fig:differential_size} shows the maximum differential visibility amplitude and phase signals that could be obtained at a spectral line with depth $x=0.85$ (as is the case for the Br$\gamma$ line of an A0V star above) as a function of the flux ratio, defined as 
\begin{align}
\max_{0 \leq \bmath{u}\cdot\bmath{\sigma} \leq 1} ||V(\bmath{u}\cdot \bmath{\sigma},f)|-|V(\bmath{u}\cdot \bmath{\sigma},xf)||
\end{align} 
and 
\begin{align}
\max_{0 \leq \bmath{u}\cdot\bmath{\sigma} \leq 1} |\arg(V(\bmath{u}\cdot \bmath{\sigma},f))-\arg(V(\bmath{u}\cdot \bmath{\sigma},xf))|
\end{align} 
While the maximum differential visibility amplitude is $\approx 8\%$, the differential visibility phase is strongly nonlinear and could be $>80\degr$ for an equal-brightness binary. Such a large signature could be used, for example, to test whether the quiescent emission from SgrA* has a contribution from a stellar component. 

\begin{figure}
\includegraphics[width=\columnwidth]{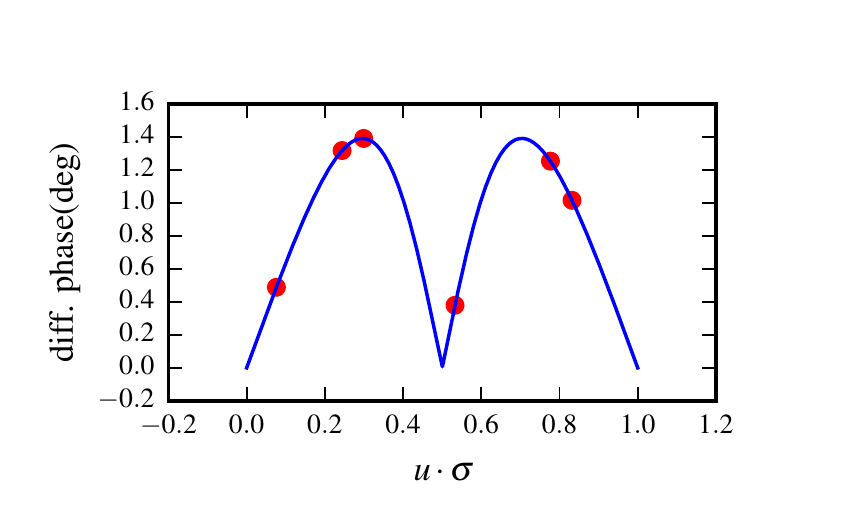} \\
 \includegraphics[width=\columnwidth]{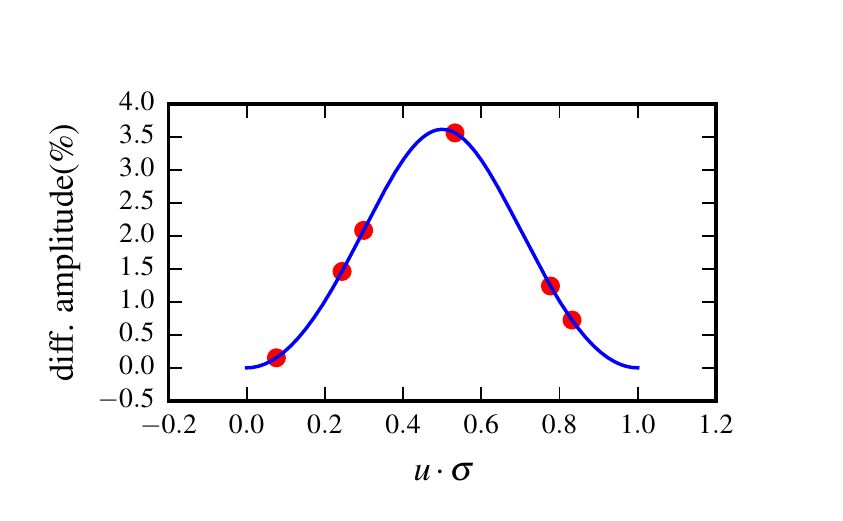}
 \caption{Differential visibility phase and amplitude across the Br$\gamma$ spectral line as a function of $\bmath{u}\cdot \bmath{\sigma}$ for the case $\Delta K=2$. The six red points show a possible sampling with the VLTI UT baselines with binary separation vector $\sigma=(10,10)$\,mas.}
 \label{fig:differential_vs_y}
\end{figure}

Here, we consider the case $\Delta K =2$ ($f=16\%$), which could correspond for e.g. to a faint star $K=18$ with a brighter component (flare or S2). We set the binary separation $\sigma=(10,10)$\,mas and onsider again the above case for the Br$\gamma$ line with depth $0.85$. Figure \ref{fig:differential_vs_y} shows the differential visibility amplitude and phase as a function of $\bmath{u}\cdot \bmath{\sigma}$, as well as the points sampled by the six VLTI UT baselines, simulated assuming $LST=18$\,h as appropriate for observing the Galactic Center. The maximum differential phase and amplitude signals are $\approx 1.5\degr$ and $4\%$, respectively. GRAVITY has already achieved a differential precision $0.2\degr$/$0.4\%$ and $0.5\degr$/$1\%$ on $K\approx6$ and $K\approx10$ sources with short integration times $\sim 1$\,h, respectively \citep{GRAVITY17}. 

In order to estimate the radial velocity precision that could be potentially achieved with such a method, we simulate differential signals on the six baselines assuming precisions $\sigma_{\phi} = 0.3\degr$ and $\sigma_{amp} = 0.6\%$, and fit all baselines simultaneously with Lorentzian profiles for the differential signatures. The resulting statistical error in redshift is $\sigma_{rv} \approx 30$\,km/s. Considering additional systematic errors of comparable order, we adopt $\sigma_{rv} = 50$\,km/s for our simulations. 

\bsp	
\label{lastpage}
\end{document}